\documentstyle[pre,aps,multicol,epsfig]{revtex}

\begin{document}

\begin{center}
{\huge Scattering of a solitary pulse on a local defect or breather} 
\vskip 
0.5truecm

Panayotis G. Kevrekidis$^{1,4}$, Boris A. Malomed$^{2,4}$, H.E. Nistazakis $
^{3}$, Dimitri J. Frantzeskakis$^{3}$, A. Saxena$^{4}$ and A.R. Bishop$^{4}$
\end{center}

$^{1}${\small Department of Mathematics and Statistics, University of
Massachusetts, Lederle Graduate Research Tower, Amherst MA 01003-4515, USA}

$^{2}${\small Department of Interdisciplinary Studies, Faculty of
Engineering, Tel Aviv University, Tel Aviv, Israel}

$^{3}${\small Department of Physics, University of Athens,
Panepistimiopolis, Zografos, Athens 15784, Greece}

$^{4}${\small Center for Nonlinear Studies and Theoretical Division, Los
Alamos National Laboratory, Los Alamos, NM 87545 USA}

\bigskip

\begin{center}
Abstract
\end{center}

A model is introduced to describe guided propagation of a linear or
nonlinear pulse which encounters a localized nonlinear defect, that may be
either static or breather-like one. The model with the static defect
directly applies to an optical pulse in a long fiber link with an inserted
additional section of a nonlinear fiber. A local breather which gives rise
to the nonlinear defect affecting the propagation of a narrow optical pulse
is possible in a molecular chain. In the case when the host waveguide is
linear, the pulse has a Gaussian shape. In that case, an immediate result of
its interaction with the nonlinear defect can be found in an exact
analytical form, amounting to transformation of the incoming Gaussian into
an infinite array of overlapping Gaussian pulses. Further evolution of the
array in the linear host medium is found numerically by means of the Fourier
transform. An important ingredient of the linear medium is the third-order
dispersion, that eventually splits the array into individual pulses. If the
host medium is nonlinear, the input pulse is naturally taken as a
fundamental soliton. The soliton is found to be much more resistant to the
action of the nonlinear defect than the Gaussian pulse in the linear host
medium, for either relative sign of the bulk and local nonlinearities. In
this case, the third-order-dispersion splits the soliton proper and
wavepackets generated by the action of the defect.

\newpage

\section{Introduction}

The interaction of traveling solitary pulses (which, in particular but not
necessarily, may be solitons, that we realize here as pulses in nonlinear
media maintaining a steady shape while propagating) with various local
imperfections or pinned dynamical excitations is a problem of fundamental
importance. For direct experimental observation, the most straightforward
case is the propagation of a pulse in an optical fiber (linear or
nonlinear), in which a strong localized nonlinear defect represents an
inserted piece of a {\it dispersion-shifted} fiber (DSF) \cite{Burtsev},
that has finite nonlinearity and negligible dispersion. This configuration,
which may be quite useful for optical telecommunications \cite{Burtsev}, is
described by the following version of the nonlinear Schr\"{o}dinger (NLS)
equation, 
\begin{equation}
iu_{z}-\frac{1}{2}\beta _{2}u_{\tau \tau }+\frac{i}{6}\beta _{3}u_{\tau \tau
\tau }+\gamma \left| u\right| ^{2}u+\Gamma \delta \left( z\right) \,\left|
u\right| ^{2}u=0,  \label{model1}
\end{equation}
where $u(z,\tau )$ is a local amplitude of the electromagnetic wave, $z$ is
the propagation coordinate, 
\begin{equation}
\tau \equiv t-z/V_{0}  \label{tau}
\end{equation}
is the so-called reduced time, $V_{0}$ being the group velocity of the
carrier wave, $\beta _{2}$ and $\beta _{3}$ are coefficients of the
second-order and third-order group-velocity dispersion (GVD) in the fiber,
and $\gamma $ is the nonlinearity coefficient of the host (system) fiber.
The strength of the localized nonlinear defect is $\Gamma \equiv \gamma _{
{\rm DSF}}L$, where $\gamma _{{\rm DSF}}$ and $L$ are the nonlinearity
coefficient and actual length of the above-mentioned finite-length DSF
inserted into the host fiber at the point $z=0$; this insertion may be
represented by the delta-function in Eq. (\ref{model1}), as the DSF's
dispersion is negligible \cite{Burtsev}. Note that the nonlinear coefficient 
$\gamma $ is always positive, as it is induced by the Kerr effect in the
optical fiber, that always has the sign corresponding to self-focusing \cite
{Agr}. On the other hand, the GVD coefficient $\beta _{2}$ may be either
positive or negative, which corresponds, respectively, to normal and
anomalous dispersion \cite{Agr}. The third-order dispersion (TOD)
coefficient $\beta _{3}$ is frequently neglected, unless $\beta _{2}$ is
very small or if the pulse is very narrow in the $\tau $-domain. However,
TOD will play an essential role in the present work. Note that $\beta _{3}$
is usually positive in optical fibers \cite{Agr}.

A qualitatively different version of the model containing the local defect
pertains to the case when it describes the interaction of the propagating
modes (pulses) with a localized defect in the form of a pinned {\em breather}
. These are spatially localized, time-periodic dynamical states, which are
ubiquitous in nonlinear physical systems ranging from quasi-one dimensional
polymers \cite{avn2} and charge-density-wave materials (e.g., metal-halogen
electronic chains \cite{avn3}) to Josephson ladders \cite{avn4}. Scattering
of a moving solitary pulse on a breather is an important process in many
physical, chemical and biological systems which combine charge, spin and
energy localization with transport of these quantities by pulses. Detailed
understanding of this type of the scattering will not only yield valuable
information on the dynamical properties of breathers, such as their mobility
and stability, but also reveal the dynamical response of materials in which
the breathers are excited. In particular, the interaction of a narrow
optical pulse with a pre-existing electron-phonon breather in a lattice
chain is a process of considerable fundamental and technological interest,
since it may enhance optical nonlinearities of the material, and thus the
efficiency of the second- and third-harmonic generation. Indeed, there is
some evidence of such an effect in conjugated polymers such as polyenes \cite
{avn1}.

Another relevant example of the breathers is provided by the rotational
dynamics of certain chemical groups in a molecular chain \cite{avn5}.
Analysis of the pulse scattering on them, in conjunction with recent
developments in ultrafast (femtosecond) spectroscopy \cite{avn6} and
chemistry \cite{avn7}, will enable an efficient use of the rich photophysics
of functional optical and electronic materials. Damage tracks in certain
mica minerals and sputtering on crystal surfaces have also been attributed
to moving breathers \cite{avn8}. In addition to mixed-valence
transition-metal complexes \cite{avn3}, there is experimental evidence for
localized breather-like states in antiferromagnetic chains \cite{avn9}, as
well as in the above-mentioned Josephson junction arrays and ladders \cite
{avn4}.

The one-dimensional (1D) model introduced and considered in this work is a
first step in a systematic study of the pulse-breather scattering, the
eventual objective being to elucidate the interaction of pulses and
breathers in 2D and 3D nonlinear dynamical lattices. However, the
formulation and consideration of the corresponding multidimensional models
is a complicated issue, which is beyond the scope of this work.

The simplest generalization of Eq. (\ref{model1}) for the case when the
local nonlinearity is induced by a small-size breather oscillating at a
frequency $\omega $ is 
\begin{equation}
iu_{z}-\frac{1}{2}\beta _{2}u_{\tau \tau }+\frac{i}{6}\beta _{3}u_{\tau \tau
\tau }+\gamma \left| u\right| ^{2}u+\Gamma \delta \left( z\right) \cos
^{2}\left( \omega t\right) \cdot \left| u\right| ^{2}u=0.  \label{model2}
\end{equation}
Equation (\ref{model2}) implies that oscillations of the breather modulate
the strength of the corresponding local nonlinear defect, but do not change
its sign, which is a natural assumption in the case of the cubic
nonlinearity. Note that, despite the difference between the time $t$ and the
reduced time $\tau $, see Eq. (\ref{tau}), it makes no difference what
definition of time is used in the argument of $\cos ^{2}$ in Eq. (\ref
{model2}), as $\tau \equiv t$ at $z=0$.

Both equations (\ref{model1}) and (\ref{model2}) conserve the norm of the
wave field (which has the physical meaning of energy in the application to
fiber optics \cite{Agr}), 
\begin{equation}
E=\int_{-\infty }^{+\infty }\left| u(\tau )\right| ^{2}d\tau .  \label{E}
\end{equation}
Additionally, the model (\ref{model1}) including the static defect conserves
the field momentum, 
\begin{equation}
P=i\int_{-\infty }^{+\infty }u_{\tau }u^{\ast }d\tau ,  \label{P}
\end{equation}
the asterisk standing for the complex conjugation.

The simplest and perhaps most fundamental version of both models (\ref
{model1}) and (\ref{model2}) is the one with all the nonlinearity
concentrated at the point $z=0$, while the bulk nonlinearity is negligible,
i.e., $\gamma =0$. It appears, rather surprisingly, that, unlike the more
technically involved case of the fully nonlinear model with $\gamma >0$,
which was studied in a part in Ref. \cite{Burtsev}, in the case $\gamma =0$
the interaction of the moving pulse with the nonlinear defect has not been
yet considered. Therefore, the detailed study of the pulse-defect
interaction in the model with $\gamma =0$ is the first objective of the
present work. Then, we will also consider the full nonlinear model with $
\gamma \neq 0$, concluding that, as a matter of fact, the most interesting
results can be obtained just in the model with the linear host medium, $
\gamma =0$.

In the linear medium, it is natural to take an incident pulse in the form of
a Gaussian, which is an eigenmode of the linear system. The problem of the
interaction of a Gaussian with the delta-like nonlinear defect can be
partially solved in an analytical form, which is done below in section 2. In
fact, the analytical result can be obtained in an advanced form for the case
of the {\it static} defect with $\omega =0$, which corresponds to Eq. (\ref
{model1}) and, as it was explained above, has direct application to fiber
optics. A basic result produced by the analytical consideration is that the
initial Gaussian pulse, passing through the nonlinear defect, generates an
infinite series of strongly overlapped Gaussian pulses. At this stage of the
analysis, TOD becomes a crucially important ingredient of the model: we show
that it lends each Gaussian pulses its own velocity, which will bring about
eventual separation of the pulses. In section 2, the velocity generated by
TOD is found in an analytical form, which yields an exact result for an
initial stage of the evolution. The splitting of the solution into an array
of pulses is considered in detail in section 3.

In section 4 we display results of the fully numerical solution for the case
when the incident Gaussian pulse interacts with the dynamical defect [the
breather, see Eq. (\ref{model2})]. In this case ($\omega \neq 0$), the
eventual pattern is much less regular, consisting of separated pulses with
random shapes, rather than Gaussian-like ones. However, if the frequency $
\omega $ is very large, one should expect that the dynamical defect
described by the last term in Eq. (\ref{model2}) may be replaced by its
averaged static counterpart. In accordance with this expectation, the
numerical computations demonstrate return to a regular pattern for very
large values of $\omega $.

In section 5 we present a solution for the most general model, which
includes the bulk nonlinearity at $z\neq 0$, i.e., $\gamma \neq 0$ in Eq. (
\ref{model2}). In this case, it is natural to take the incident pulse as a 
{\em soliton} of the corresponding homogeneous NLS equation, rather than a
Gaussian pulse, and the interaction may only be simulated numerically.
Results of simulations of the interaction of the soliton with the defect
turn out to be quite different from those in the case when the host medium
was linear: the soliton is found to be much more resistant to the action of
the local nonlinear defect than the Gaussian pulse in the linear medium.
Unless the defect is extremely strong, the pulse remains virtually intact,
the TOD term separating it from small-amplitude wave packets generated by
the local defect. However, in the absence of the TOD term, the effect of the
perturbation may accumulate and destroy the soliton. Thus, the third-order
dispersion plays a crucially important role in {\em both} versions of the
present model, i.e., for the linear and nonlinear host medium.

If the defect is very strong, it may split the soliton into two. The results
obtained for the model with the nonlinear host medium are not sensitive to
the value of the defect's intrinsic frequency $\omega $, being essentially
the same for the static defect and its dynamical counterpart. Moreover, the
soliton is found to be stable against the action of the local defect
irrespective of the relative sign of the bulk and local nonlinearities.
These results, obtained for the case when the input signal is a soliton,
rather than a linear pulse, is another manifestation of the well-known
general principle, according to which solitons are very robust eigenmodes of
nonlinear media.

\section{Analytical consideration of the interaction between a Gaussian
pulse and a nonlinear defect.}

\subsection{The pulse}

We start the consideration with the simplest version of the model, viz.,
Eqs. (\ref{model1}) or (\ref{model2}) with $\gamma =\beta _{3}=0$. At $z\neq
0$, we thus have a linear Schr\"{o}dinger equation, which gives rise to the
well-known exact solution in the form of a Gaussian pulse (which is termed a
coherent state in quantum mechanics), 
\begin{equation}
u(z,\tau )=\frac{A_{0}}{\sqrt{1-2i\beta _{2}cz}}\,\exp \left[ -\frac{c\left(
1+2i\beta _{2}cz\right) }{1+\left( 2\beta _{2}cz\right) ^{2}}\cdot \tau ^{2} 
\right] \,,  \label{Gauss}
\end{equation}
where $c>0$ and $A_{0}$ are arbitrary real constants. The constant $c$
determines the width of the Gaussian pulse and its {\it chirp} (i.e., the
imaginary part of the coefficient in front of $\tau ^{2}$ in the argument of
the exponential, which shows a slope of the local frequency across the pulse 
\cite{Agr}).

Note that the solution (\ref{Gauss}) has zero velocity in the present
reference frame [which is defined by Eq. (\ref{tau})]. An exact solution for
a moving pulse can be generated from Eq. (\ref{Gauss}) by the action of a 
{\it boost} (Galilean transform) 
\begin{equation}
u(z,t)\rightarrow u\left( z,\tau -sz\right) \exp \left( \frac{is^{2}}{2\beta
_{2}}z-\frac{is}{\beta _{2}}\tau \right) ,  \label{s}
\end{equation}
where $s$ is a real velocity-shift parameter. It is important to notice
that, for the boosted pulse, the values of its velocity, momentum (\ref{P})
and energy (\ref{E}) are related, irrespective of the particular form of the
pulse, in a simple way, 
\begin{equation}
s=\beta _{2}\frac{P_{{\rm pulse}}}{E_{{\rm pulse}}}\,.  \label{boosted}
\end{equation}

\subsection{Pulse acceleration by the third-order dispersion and numerical
verification}

Before proceeding to the consideration of the passage of the pulse (\ref
{Gauss}) through the nonlinear defect, we need to understand how a free
pulse will move under the action of the TOD term in Eqs. (\ref{model1}) or (
\ref{model2}). The consideration of this issue will help to understand how
an array of pulses splits in the presence of the TOD term.To
this end, a solution is sought for by means of the Fourier transform, which
yields the following integral representation for it: 
\begin{equation}
u(z,\tau )=\frac{A_{0}}{2\sqrt{\pi c}}\int_{-\infty }^{+\infty }\exp \left[ -
\frac{\omega ^{2}}{4c}-i\omega \tau +i\left( \frac{1}{2}\beta _{2}\omega
^{2}-\frac{1}{6}\beta _{3}\omega ^{3}\right) z\right] d\omega
\label{Fourier}
\end{equation}
[setting $z=0$, this expression goes over into Eq. (\ref{Gauss}) taken at $
z=0$]. 
In the limit of large values of $z$, the integral (\ref{Fourier}) is
dominated by a contribution from a vicinity of the stationary-phase point,
which is 
\begin{equation}
\omega _{0}\approx \frac{\tau }{\beta _{2}z}-\frac{i\tau }{2c\beta
_{2}^{2}z^{2}}+\frac{\beta _{3}\tau ^{2}}{2\beta _{2}^{3}z^{2}}.
\label{omega0}
\end{equation}
In particular, a contribution of the stationary-phase point (\ref{omega0})
to the phase of the pulse is 
\begin{equation}
\phi (z,\tau )=\frac{\tau ^{2}}{2\beta _{2}z}-\frac{\beta _{3}\tau ^{3}}{
6\beta _{2}^{3}z^{2}}.  \label{phi}
\end{equation}
Direct simulations show that, under the action of TOD, the initial Gaussian
pulse is gradually destroyed, generating a long ``tail'', while a part of
the wave packet may still be interpreted as a surviving pulse. Using the
expression (\ref{phi}), one can find a relation between the momentum [see
Eq. (\ref{P})] and energy of the pulse-like part of the wave packet, 
\begin{equation}
P_{{\rm pulse}}=\frac{\beta _{3}c}{2\beta _{2}}E_{{\rm pulse}}.  \label{PM}
\end{equation}
Note that, as the {\em net} momentum of the wave field must be conserved, a
``recoil'' momentum $-P_{{\rm pulse}}$ is carried away by the
above-mentioned tail.
The comparison of Eqs. (\ref{boosted}) and (\ref{PM}) yields an analytical
prediction for the asymptotic value of the velocity shift acquired by the
surviving pulse under the action of the TOD term in the limit $z\rightarrow
\infty $: 
\begin{equation}
s=(1/2)\beta _{3}c.  \label{sc}
\end{equation}
To this end, a natural definition of the position of the wave-packet's
center of mass is adopted, 
\begin{equation}
\tau _{c}\equiv E^{-1}\int_{-\infty }^{+\infty }\tau \left| u(\tau )\right|
^{2}d\tau ,  \label{tau_c}
\end{equation}
where $E$ is the net energy defined by Eq. (\ref{E}), and the derivative $
d\tau _{c}/dz$ may be regarded as a velocity of the pulse.

In the case of the free propagation, one can derive an exact relation 
\begin{equation}
\frac{d\tau _{c}}{dz}=\beta _{2}\frac{P}{E}-\frac{1}{2}\beta
_{3}E^{-1}\int_{-\infty }^{+\infty }\left| u_{\tau }\right| ^{2}d\tau ,
\label{velocity}
\end{equation}
where $P$ and $E$ are the net momentum and energy of the wave field.
Comparing this to Eq. (\ref{boosted}), we conclude that, in the absence of
TOD, the velocity (\ref{velocity}) is identical to the boost parameter (in
particular, $d\tau _{c}/dz=0$ if $P=0$); however, this identity is broken by
TOD, that is why acceleration of the pulse by TOD is observed. If the TOD
term is treated as a small perturbation, one can analytically calculate the
second term on the right-hand side of Eq. (\ref{velocity}), using the
expression (\ref{Gauss}), which is an exact solution in the absence of TOD.
As a result, we find the velocity of the pulse in the case when the net
field momentum vanishes, $P=0$, which is true for the initial Gaussian pulse
(\ref{Gauss}): 
\begin{equation}
\frac{d\tau _{c}}{dz}=-\frac{1}{3}\beta _{3}c  \label{motion}
\end{equation}
[note that this expression does not contains neither $z$ nor $\beta _{2}$,
despite the fact the solution (\ref{Gauss}), used for the calculation of the
right-hand side of Eq. (\ref{motion}), does depend on $z$ and $\beta _{2}$].
The expression (\ref{motion}) shows that the velocity lent to the pulse by
TOD linearly depends on the pulse's parameter $c$, hence initially
overlapping pulses with different values of $c$ [see Eq. (\ref
{expanded_output}) below] are expected to separate under the action of TOD.

The analytical result (\ref{expanded_output}) was checked against direct
numerical simulations of the linear version of Eq. (\ref{model1}). For
instance, in the case $\beta _{2}=1$, $\beta _{3}=0.1$, and $c=1$, it was
found that the pulse was indeed moving at a constant velocity, the value of
which exactly coincided with that given by Eq. (\ref{motion}), in the
interval $0<z<25$. At larger values of $z$, the absolute value of the
velocity decreases, which can be explained by the fact that the TOD term
essentially alters the shape of the pulse at that later stage of the
evolution.

\subsection{Passage of the pulse through the nonlinear defect}

Our next aim is to consider transformation of the Gaussian pulse passing the
nonlinear defect. Obviously, in an infinitesimal vicinity of the defect (as $
|z|\rightarrow 0$), only the first and last terms should be kept in Eqs. (
\ref{model1}) and (\ref{model2}), which yields a simplified equation, 
\begin{equation}
\frac{\partial u}{\partial z}=i\Gamma \delta \left( z\right) \cos ^{2}\left(
\omega \tau \right) \cdot \left| u\right| ^{2}u.  \label{ODE}
\end{equation}
To solve Eq. (\ref{ODE}), we represent the solution as $u(z)\equiv
a(z)\,\exp \left[ i\phi (z)\right] $, with real amplitude $a$ and phase $
\phi $. Substituting this into Eq. (\ref{ODE}), one immediately finds that $
\partial a/\partial z=0$, and 
\[
\frac{\partial \phi }{\partial z}=\Gamma \delta \left( z\right) \cos
^{2}\left( \omega \tau \right) \cdot a^{2}. 
\]
A solution to the latter equation is obvious, 
\begin{equation}
\phi (z=+0,\tau )-\phi (z=-0,\tau )=\Gamma a^{2}\left( \tau \right) \cdot
\cos ^{2}\left( \omega \tau \right) ,  \label{phase}
\end{equation}
where we take into regard that $a$ may be a function of $\tau $.

Thus, we take the input pulse at the point $z=-0$ in the general form [cf.
the expression (\ref{Gauss})], 
\begin{equation}
u(z=-0)=A_{0}\,\exp \left[ -\left( c_{0}+ib_{0}\right) \tau ^{2}\right] ,\,
\label{input}
\end{equation}
where $c_{0}>0$ determines the initial width of the pulse, and $b_{0}$ is
its initial chirp. The substitution of the expression (\ref{input}) into Eq.
(\ref{phase}) yields the form of the pulse appearing after the passage of
the nonlinear defect: 
\begin{equation}
u(z=+0,\tau )=A_{0}\,\exp \left[ -\left( c_{0}+ib_{0}\right) \tau
^{2}+i\Gamma A_{0}^{2}\exp \left( -2c_{0}\tau ^{2}\right) \cdot \cos
^{2}\left( \omega \tau \right) \right] .  \label{output}
\end{equation}

Further analytical consideration for the general case of the dynamical
defect (breather), with $\omega \neq 0$, is extremely cumbersome. Therefore,
in the rest of this section and in the next one, we focus on the static
case, $\omega =0$. The aim will be to realize a result of the further
evolution of the transformed pulse (\ref{output}), governed by the linear
equation (\ref{model1}) Including the TOD term) at $z>0$. To this end, we
notice that the expression (\ref{output}) can be expanded into an infinite
series: 
\begin{equation}
u(z=+0)=A_{0}\exp \left( -ib_{0}\tau ^{2}\right) \,\sum_{n=0}^{+\infty
}\left( n!\right) ^{-1}\left( i\Gamma A_{0}^{2}\right) ^{n}\exp \left[
-\left( 1+2n\right) c_{0}\tau ^{2}\right] .  \label{expanded_output}
\end{equation}
Comparing it to the exact fundamental-pulse solution (\ref{Gauss}), we
conclude that the expression (\ref{expanded_output}), if considered as an
initial condition to the linear equation (\ref{model1}) with $\gamma =\beta
_{3}=0$, gives rise to a superposition of an infinite number of Gaussians
with the values of the width constant $c_{n}=\left( 1+2n\right) c_{0}$, and
with the common initial value $b_{0}$ of the chirp. The evolution of the
wave packet (\ref{expanded_output}) can be presented in a relatively simple
form in the case when the initial chirp is absent, $b_{0}=0$ (and TOD is
neglected, $\beta _{3}=0$): 
\begin{equation}
u(z,\tau )=A_{0}\,\exp \left( -ib_{0}\tau ^{2}\right) \sum_{n=0}^{+\infty } 
\frac{\left( i\Gamma A_{0}^{2}\right) ^{n}}{n!\sqrt{1-2i\beta
_{2}c_{0}\left( 1+2n\right) z}}\exp \left[ -\frac{c_{0}\left( 1+2i\beta
_{2}c_{0}\left( 1+2n\right) z\right) }{1+\left( 1+2n\right) ^{2}\left(
2\beta _{2}c_{0}z\right) ^{2}}\tau ^{2}\right]  \label{Gaussians}
\end{equation}
[in the case $\omega \neq 0$, Eq. (\ref{output}) shows that the Gaussian in
each term of the initial \ \ series (\ref{expanded_output}) with $n\neq 0$
is additionally multiplied by $\left[ \cos (\omega t)\right] ^{2n}$, which
will make the subsequent result much more complex than that given by Eq. (
\ref{Gaussians})].

Thus, Eq. (\ref{Gaussians}) gives an exact solution to the nonlinear model
equation (\ref{model1}) in the case when the incident pulse has no chirp and 
$\gamma =\beta _{3}=0$. However, while taking the input pulse to be
chirpless, and disregarding the bulk nonlinearity ($\gamma =0$) are quite
acceptable assumptions, the TOD term may {\em not} be neglected, as, without
this term, the exact solution (\ref{Gaussians}) remains strongly degenerate.
Indeed, centers of all the Gaussian pulses, the superposition of which
constitutes this solution, exactly coincide, staying at $\tau =0$ [note that
this degeneracy is not lifted if the input pulse has nonzero chirp, nor if
the bulk nonlinearity ($\gamma \neq 0$) is added]. On the other hand, Eq. (
\ref{motion}) shows that the TOD term will lend each pulse its own velocity,
depending on the initial width parameter $c$ of the pulse. Obviously, this
will eventually split the superposition (\ref{Gaussians}) of the overlapping
Gaussians into an array of separating pulses.

However, the analytical prediction (\ref{motion}) for the TOD-induced
velocity shift of each pulse is only valid for finite values of $z$.
Therefore, to find an actual shape of the evolving wave train, it is
necessary to solve numerically the linear Schr\"{o}dinger equation with the
TOD term, 
\begin{equation}
iu_{z}-\frac{1}{2}\beta _{2}u_{\tau \tau }+\frac{i}{6}\beta _{3}u_{\tau \tau
\tau }=0,  \label{TOD}
\end{equation}
with the initial condition in the form (\ref{output}). Results generated by
Eq. (\ref{TOD}) are presented in the next section.

\section{Generation of the wave train by the third-order dispersion in the
case of the static nonlinear defect}

As Eq. (\ref{TOD}) is linear and has constant coefficients, it can be solved
by Fourier transform. Thus, the numerical part of the solution amounts to
the computation of the Fourier transform 
\[
u_{0}(\omega )\equiv \int_{-\infty }^{+\infty }\exp (i\omega \tau
)u(z=+0,\tau )d\tau 
\]
for the initial configuration (\ref{output}), and subsequent computation of
the inverse Fourier transform 
\[
u(z,\tau )\equiv (2\pi )^{-1}\int_{-\infty }^{+\infty }\exp (-i\omega \tau
)u(z,\omega )d\tau , 
\]
where, as it immediately follows from Eq. (\ref{TOD}), 
\[
u(z,\omega )=u_{0}(\omega )\,\exp \left[ i\left( \frac{1}{2}\beta _{2}\omega
^{2}z-\frac{1}{6}\beta _{3}\omega ^{3}z\right) \right] \,. 
\]

In Fig. 1 we display the profiles of $|u(\tau )|$, computed at the points $
z=10,\,20,\,30,$ and $40$ for a case when the static nonlinear defect is
weak, having $\Gamma =0.1$. The figure demonstrates that, in accordance with
the analysis presented in the previous section, the initial wave packet
tends to split into an array of regular Gaussian-like pulses. The splitting
actually takes place for larger values of the nonlinear-defect's strength,
as is shown in Fig. 2, which pertains to $\Gamma =1$. For the same case, the
eventual shape of the pulse array is shown in more detail in Fig. 3.

The result for a still stronger nonlinear defect, with $\Gamma =10$, is
displayed in Fig. 4. In this case, the splitting into pulses takes a violent
character, with appearance of huge gradients and formation of a very sharp
front, the latter effect being accounted for by the interplay of the strong
local nonlinearity and TOD. These results may be explained by the fact that,
as follows from Eq. (\ref{expanded_output}), the number $n_{\max }$ of the
largest-amplitude pulse in the series grows $\sim \Gamma $ with increase of $
\Gamma $, hence the width $W$ of an individual pulse decreases $\sim 1/\sqrt{
\Gamma }$, which gives rise to the large gradients revealed by the
subsequent evolution. We also note that, although the TOD coefficient $
\beta_{3}$ remains small, the size of the TOD term in Eqs. (\ref{model1})
and (\ref{model2}) grows as $1/W^{3}$ with the decrease of the pulse's
width. This explains the formation of the abrupt front under the action of
the asymmetric TOD term.

\section{Generation of the pulse array by a localized breather}

An example of the splitting of the initial Gaussian as a result of its
interaction with the dynamical defect (with the same strength $\Gamma =1$ as
in the case shown in Fig. 2, and the frequency $\omega =1/2$) is shown in
Fig. 5. The splitting is displayed in more detail by blowups collected in
Fig. 6.

Comparison of Figs. 2 and 5 shows that the static defect and the dynamical
one with a moderate value of the frequency produce quite similar results.
Taking larger values of the frequency strongly changes the situation: as is
shown in Fig. 7 and in the accompanying blowup (Fig. 8), the same value of
the defect strength as in the cases displayed in Figs. 2 and 5, i.e., $
\Gamma =1$, but combined with $\omega =10$, gives rise to an essentially
more disordered pattern. Note that this pattern is disordered in a way
essentially different from that observed as a result of the action of a
strong static defect ($\omega =0$) , cf. Fig. 4. In particular, strong
asymmetry of the pattern and sharp fronts are not found in the present case,
and the local gradients are not as huge as in Fig. 4. The absence of those
features in the present case is easy to understand, as they may only be
generated by a large value of $\Gamma $, as explained above.

Keeping to increase $\omega $ at a fixed value of $\Gamma $, we have
concluded that the situation shown in Figs. 7 and 8 is rather similar to
that observed at $\Gamma =1$ and $\omega =100$ (not shown here). On the
other hand, it is obvious that, if the frequency is extremely large, one
should be able to replace $\cos ^{2}(\omega t)$ in Eq. (\ref{model2}) by its
average value $1/2$, thus reverting to Eq. (\ref{model1}) for the static
nonlinear defect. To verify this argument, in Fig. 9 we present the results
of the numerical computations for $\Gamma =1$ and $\omega =1000$, which are
supported by the blowup shown in Fig. 10. From these pictures, it is evident
that, in the case of extremely large $\omega $, the situation is indeed
nearly the same as in the case of the static nonlinear defect, cf. Figs. 2
and 3.

\section{Transformation of a soliton by the local defect in the nonlinear
host medium}

In all the cases considered above, the input pulse was taken in the form of
a Gaussian, since the host medium in which this pulse propagated before the
collision with the nonlinear defect was linear, the Gaussian being its
eigenmode. The situation is completely different in the case when the host
waveguide is itself nonlinear, as in that case the incoming pulse must be a $
{\rm sech}$ soliton \cite{Burtsev}. The action of the localized nonlinearity
on the soliton is described by the same general expression (\ref{phase}) as
above; however, the result cannot be interpreted in such a straightforward
manner (even if $\omega =0$) as it was done above for the case of the input
Gaussian pulse, see Eq. (\ref{expanded_output}). In fact, in the case of the
nonlinear host medium all the analysis following the use of expression (\ref
{phase}) for the defect-induced phase change of the input soliton can only
be performed numerically.

As is commonly known, in the homogeneous part of the nonlinear model (\ref
{model1}) or (\ref{model2}) (at $z\neq 0$), the soliton may only exist if
the second-order dispersion is anomalous while the nonlinearity is
self-focusing, or vice versa \cite{Agr}, i.e., only if $\beta _{2}\gamma <0$
. In the simulations, we fixed $\beta _{2}\equiv +1$, and took different
negative values of $\gamma $, see below. The input soliton was always\ taken
as a fundamental one with a fixed width corresponding to these values of the
parameters: 
\begin{equation}
u\left( z=-0,\tau \right) =|\gamma |^{-1/2}{\rm sech\,}\tau \,.
\label{soliton}
\end{equation}

If the TOD term is neglected ($\beta _{3}=0$), the nonlinear defect produces
a strong perturbation around the soliton. However, the TOD term helps to
separate the soliton and the perturbation, as is illustrated by Fig. 11 for
the case $\beta _{3}=0.3$. This figure shows a large $\tau $-domain, in
order to demonstrate the evolution over a large propagation distance; as is
seen, the perturbation spreads out indefinitely, while the soliton remains
essentially intact.

The comparison of Fig. 11 with Fig. 2, which shows the interaction of the
Gaussian pulse in the linear host medium with the local nonlinear defect
that has the same strength, $\Gamma =1$, suggests a conclusion that the
soliton in the nonlinear host medium is much more resistant to the action of
the nonlinear defect than the Gaussian pulse in the linear host medium. In
fact, this conclusion is strongly supported by many other simulations with
different values of the parameters.

Another difference of the present case from that for the Gaussian pulse in
the linear host medium is that the result of the action of the nonlinear
defect on the soliton is less sensitive to the defect's intrinsic frequency.
For instance, if the defect is dynamical with $\omega =1$, while the other
parameters take the same values as in the case shown in Fig. 11, the
evolution (not shown here) seems nearly the same as in Fig. 11, with a
difference that the generated perturbation takes a somewhat larger portion
of energy from the soliton (as a result, the soliton reappears after the
collision with the amplitude $\approx 0.8$, to be compared with the
amplitude $\approx 1$ in Fig. 11). Nevertheless, in this case too, the
perturbation remains rather small and it does not strongly affect the
soliton.

With further increase of the dynamical defect's frequency, the perturbation
again becomes smaller, and the general picture is reverting to that
corresponding to the static defect with $\omega =0$. This trend can be
easily explained by the self-averaging of the dynamical defect similar to
that observed above in the case of the linear host medium. However, unlike
that case, where the self-averaging was evident only for extremely large
values of the frequency, $\omega \sim 1000$ (see Fig. 9), in the present
case simulations (not shown here) clearly show that $\omega =10$ is
sufficient for the self-averaging to manifest itself.

The above results were obtained for the cases with $\Gamma >0$ and $\gamma
<0 $, which implies that the localized nonlinearity is self-defocusing if
the bulk nonlinearity is self-focusing, or vice versa. It is also
interesting to consider the case when both nonlinearities have the same
sign, i.e., $\Gamma <0$. An example is displayed in Fig. 12 for $\Gamma =-1$
(this example pertains to the dynamical defect with $\omega =1$, but the
results are virtually the same for the static defect with $\omega =0$). In
this case, the action of the defect produces a weaker effect, as the soliton
reappears after the interaction with the amplitude $\approx 1$, to be
compared with the above-mentioned value $\approx 0.8$ of the soliton's
amplitude found for the same values of parameters but $\Gamma =+1$. This
difference is quite natural, as the local defect which has the same sign of
the nonlinearity as in the host medium is acting to compress the soliton
stronger, while the local defect with the opposite sign of $\Gamma $ was
acting to stretch the soliton, which produces a more destructive effect,
helping some radiation (wave packets) to escape from the weakened soliton
pulse.

Taking a much stronger nonlinear defect (with $\Gamma =11$, see Fig. 13), we
conclude that, naturally, the soliton loses a larger part of its energy to
the generation of the perturbation separating from it. Nevertheless, the
soliton survives even in this case (cf. Fig. 7, which shows a complete and
fast chaotization of the wave field in the case of the collision of the
Gaussian pulse with a strong defect, having $\Gamma =10$, in the linear host
medium). In the cases when $\Gamma $ is large but negative (not shown here),
when the nonlinearity of the defect has the same sign as in the host medium,
the soliton's losses are smaller than in the case displayed in Fig. 13,
which is quite similar to what is shown in Fig. 12 for a moderately strong
defect with $\Gamma <0$.

\section{Conclusion}

We have introduced a model describing collision of a pulse in a linear or
nonlinear waveguide with a strong nonlinear local defect, that may be either
static or breather-like. The model with the static defect should directly
apply to an optical pulse in a long fiber-optic link with an inserted
section of a nonlinear (dispersion-shifted) fiber of an arbitrary length. On
the other hand, a local breather, which gives rise to the nonlinear defect
affecting the propagation of narrow optical pulses, may be realized in a
molecular chain with electron-phonon coupling.

In the case when the host waveguide is linear, the pulse was naturally taken
as a Gaussian. A result of its interaction with the nonlinear defect was
found analytically, amounting to its transformation into a ``lump''
consisting of an infinite number of overlapping Gaussian pulses. Further
evolution of the lump in the linear medium is generated by the corresponding
Fourier transform. An important ingredient of the medium is the third-order
dispersion, that splits the lump into an array of individual pulses; the
velocity shift lent to an initially Gaussian pulse by the TOD term was found
in an analytical form. The influence of the intrinsic frequency, in the case
when the defect is a breather, was also investigated.

The full numerical solution for the model in which the host medium is
nonlinear, and the input pulse is taken as a fundamental soliton shaped by
this medium, produces a result quite different from that for the linear host
medium: the soliton is much more resistant to the action of the local
nonlinearity. If the local defect is not very strong, the soliton remains
essentially intact, the third-order-dispersion separating the soliton and
small wave packets generated by the collision. Even if the local
nonlinearity is very strong, the soliton survives, losing a limited part of
its energy.

Beyond straightforward application of the results reported above to
fiber-optic links, the model introduced in this work, and current
developments in experimental techniques, such as ultrafast spectroscopy \cite
{avn6,avn7} of optical and electronic materials \cite{avn3}, the results
presented here may help to understand the dynamics of scattering of pulses
on static and breather-like defects in many physical systems, e.g.,
conjugated polymers \cite{avn1,kress,kivshar,kopa}, Josephson ladders \cite
{avn4}, and coupled electron-vibron lattice systems \cite{hennig}.

\section*{Acknowledgements}

B.A.M. appreciates hospitality of the Center for Nonlinear Studies at the
Los Alamos National Laboratory and of the Department of Physics at the
University of Athens (Greece). Work at Los Alamos was supported by the U.S.
Department of Energy.

\newpage

\section*{Figure Captions}

Fig. 1. The evolution of the field $|u(\tau )|$ after the interaction of the
input Gaussian pulse with the nonlinear defect, in the case $\omega =\gamma
=0$, $\beta _{2}=1$, $\beta_{3}=0.1$, and $\Gamma =0.1$.

Fig. 2. The same as in Fig. 1 for a stronger nonlinear defect, with $\Gamma
=1$.

Fig. 3. Blowups of the field pattern from the last panel in Fig. 2, clearly
showing the formation of an array of separated pulses.

Fig. 4. The same as in Fig. 3 for a very strong static nonlinear defect,
with $\Gamma =10$.

Fig. 5. The evolution of the field $|u(\tau )|$ after the interaction of the
input Gaussian with the nonlinear dynamical defect, in the case $\omega =1/2$
, $\gamma =0$, $\beta _{2}=1$, $\beta _{3}=0.1$, and $\Gamma =1$. Except for 
$\omega $, these parameters are the same as in the case shown in Fig. 2.

Fig. 6. Blowups of segments of the panels from Fig. 5, showing the gradual
splitting of the wave packet into an array of pulses.

Fig. 7. The same as in Fig. 5, but with $\omega =10$.

Fig. 8. The same as in Fig. 6, but with $\omega =10$.

Fig. 9. The same as in Fig. 7, but with $\omega =1000$.

Fig. 10. The same as in Fig. 6, but with $\omega =1000$.

Fig. 11. The evolution of the field $|u(\tau)|$ over a very large
propagation distance in a large temporal domain after the interaction of the
input soliton (\ref{soliton}) with the nonlinear static defect, in the case $
\gamma =-\beta _{2}=-1$, $\beta _{3}=0.3$, and $\Gamma =1$, $\omega =0$. It
is evident that the local defect initially creates a large perturbation
around the soliton; however, the perturbation is diffused away under the
combined action of the second- and third-order dispersions, and a robust
soliton reappears.

Fig. 12. The same as in Fig. 11, but with $\omega =1$ and $\Gamma =-1 $.

Fig. 13. The same as in Fig. 12, but with $\beta_3=0.5$ and $\Gamma =+11$.


\newpage

\begin{references}
\bibitem{Burtsev}  S. Burtsev, D. J. Kaup, and B. A. Malomed, 
{\it Interactions of solitons with a strong inhomogeneity in a nonlinear 
optical fiber}, 
Phys. Rev. E {\bf 52}, 4474 (1995).

\bibitem{Agr}  G. P. Agrawal, {\it Nonlinear Fiber Optics} (Academic Press:
Boston, 1995).

\bibitem{kress}  J. D. Kress, A. Saxena, A. R. Bishop, and R. L. Martin,
Phys. Rev. B 58, 6161 (1998).

\bibitem{kivshar}  A. Saxena, Y. S. Kivshar, and A. R. Bishop, Synth. Met. 
{\bf 116}, 45 (2001).

\bibitem{kopa}  G. Kopidakis and S. Aubry, Physica B 296, 237 (2001).

\bibitem{avn2}  S. R. Phillpot, A. R. Bishop, and B. Horovitz, Phys. Rev. B
40, 1839 (1989).

\bibitem{avn3}  B. I. Swanson, J. A. Brozik, S. P. Love, G. F. Strouse, A.
P. Shreve, A. R. Bishop, W. Z. Wang, and M. I. Salkola, Phys. Rev. Lett. 82,
3288 (1999).

\bibitem{avn4}  P. Binder, D. Abraimov, A. V. Ustinov, S. Flach, and Y.
Zolotaryuk, Phys. Rev. Lett. {\bf 84}, 745 (2000).

\bibitem{avn1}  H. W. Streitwolf, Phys. Rev. B 58, 14356 (1998).

\bibitem{avn5}  F. Fillaux, C. J. Carlile, and G. J. Kearley, Phys. Rev. B
58, 11416 (1998).

\bibitem{avn6}  A. H. Zewail, Adv. Chem. Phys. 101, 892 (1997).

\bibitem{avn7}  A. H. Zewail, J. Phys. Chem. 100, 12701 (1996).

\bibitem{avn8}  J. L. Marin, J. C. Eilbeck, and F. M. Russell, in {\it 
Nonlinear Science at the Dawn of the 21st Century}, eds. P. L. Christiansen
and M. P. Sorensen (Springer, Berlin, 2001).

\bibitem{avn9}  U. T. Schwarz, L. Q. English, and A. J. Sievers, Phys. Rev.
Lett. 83, 223 (1999).

\bibitem{hennig}  D. Hennig, Phys. Rev. E {\bf 62}, 2846 (2000).
\end{references}
\end{document}